\documentclass[rmp,aps,twocolumn]{revtex4}
\newcommand{\la}{\langle}
\newcommand{\ra}{\rangle}
\newcommand{\ua}{\uparrow}
\newcommand{\da}{\downarrow}

\newcommand{\e}{\epsilon}

\usepackage{dcolumn}
\usepackage{amsmath}
\usepackage{color}
\usepackage{graphicx}
\usepackage{amssymb}

\begin{document}

\title{Emergence of magnetism in graphene materials and nanostructures}

\author{Oleg V. Yazyev}
\email[Electronic address: ]{yazyev@civet.berkeley.edu}
\affiliation{Department of Physics, University of California, Berkeley, California 94720, USA}
\affiliation{Materials Sciences Division, Lawrence Berkeley National Laboratory, Berkeley, California 94720, USA}

\date{\today{}}

\begin{abstract}
Magnetic materials and nanostructures based on carbon offer unique 
opportunities for future technological applications such as spintronics. 
This article reviews graphene-derived systems in which magnetic 
correlations emerge as a result of reduced dimensions, disorder and other 
possible scenarios. In particular, zero-dimensional graphene nanofragments, 
one-dimensional graphene nanoribbons, and defect-induced magnetism in graphene 
and graphite are covered. Possible physical mechanisms of the emergence of 
magnetism in these systems are illustrated with the help of computational examples
based on simple model Hamiltonians. In addition, this review covers 
spin transport properties, proposed designs of graphene-based spintronic 
devices, magnetic ordering at finite temperatures as well as the most recent 
experimental achievements.
\end{abstract}


\maketitle
\tableofcontents{}

\section{Introduction}
\label{intro}

Magnetic materials are essential for modern technology. All presently used magnetic
materials involve the elements belonging to either the $d$- or the $f$-block of 
the periodic table. For instance, among the periodic table elements only the
late transition metals Fe, Co and Ni, are ferromagnets at room temperature. 
Magnetic ordering in these transition metals originate from the partially 
filled $d$-electron bands. However, magnetism is not common for 
the light $p$-block elements belonging to the second period of the periodic table,
even despite the fact that carbon is able to form very diverse and complex 
molecular structures. In principle, such materials may possess a number 
of attractive properties such as low density, biocompatibility, plasticity, and 
many others, which stimulates the search for light-element based magnetism \cite{Makarova06a}.

\begin{figure}
\includegraphics[width=8.5cm]{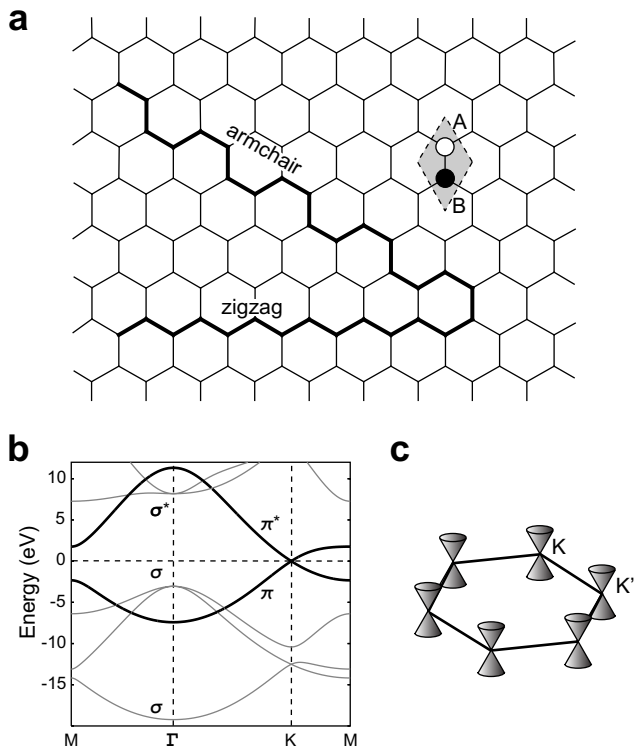}
\caption{\label{fig1} 
(a) Two-dimensional crystalline lattice of graphene. The shaded area denotes 
the unit cell of graphene containing two carbon atoms which belong
to the two sublattices of graphene, $A$ (empty circle) and $B$ 
(filled circle). The two high-symmetry directions of graphene lattice, 
armchair and zigzag, are highlighted. (b) Band structure of graphene obtained 
by means of first-principles calculations. The bands are labeled according to their 
symmetry. The $\pi$-symmetry bands responsible for the low-energy electronic
properties of graphene are highlighted. The zero energy corresponds to the 
Fermi level. (c) The low-energy part of the band structure of graphene 
involves two inequivalent `Dirac cone' features in the corners (points
$K$ and $K'$) of the hexagonal Brillouin zone.}
\end{figure}

The field of light-element magnetism and, in particular, of carbon-based 
magnetism is currently gaining increasing importance because of the following
two reasons. Firstly, the field of carbon-based magnetism has always been
a very controversial area of research which suffered from the poor reproducibility
of experimental results. However, the situation seems to be improved over
the last few years. Several examples of magnetism in carbon-based materials
continue to be reliably reproduced by different research groups.
The second driving force is the first isolation of {\it graphene}, a truly 
two-dimensional form of carbon which has attracted enormous attention in 
science and technology \cite{Novoselov04}. Graphene has a fairly 
simple honeycomb atomic structure, but rather unique electronic 
structure with linear band dispersion at the Fermi level (see Fig.~\ref{fig1})
which is largely responsible for many novel physical phenomena observed
in this material (for review see \cite{Katsnelson07,Geim07,CastroNeto09}). 
Importantly, graphene can also be considered as a unifying concept for understanding 
a broad class of $sp^2$ carbon materials which includes polycyclic aromatic molecules, 
fullerenes, carbon nanotubes, and graphite as well as their further modifications obtained 
by patterning, chemical treatment, implantation of defects, impurities, etc.  

While ideal graphene is nonmagnetic itself, many its derivative materials
and nanostructures, both realized in practice and considered theoretically, show 
various scenarios of magnetism. The magnetic graphene nanostructures are 
particularly promising for applications in the field of {\it spintronics}, a very 
probable future step in the evolution of electronics industry, which promises 
information storage, processing and communication at faster speeds and lower 
energy consumption \cite{Wolf01,Chappert07,Awschalom07,Fert08}. 
While traditional electronics exploits only the charge 
of electron, spintronics will also make use of its spin degree of freedom. 
For the field of spintronics graphene can offer a possibility of tuning its 
spin transport properties by means of various applied stimuli. For instance, it was 
suggested that half-metallicity of zigzag graphene nanoribbons can be 
triggered by external electric fields \cite{Son06b}. If realized in practice,
this would allow for efficient electric control of spin transport, a very desirable 
effect which is hard to achieve using other materials. In addition,
materials based $sp$-elements are expected to have high magnitudes of the 
spin-wave stiffness \cite{Edwards06} and, thus, nanostructures made of these 
elements would possess higher Curie temperatures or spin correlation lengths \cite{Yazyev08a}. 
Materials based on light elements also display weak spin-orbit and hyperfine couplings 
which are the main channels of relaxation and decoherence of electron spins 
\cite{Trauzettel07, Yazyev08b, Fischer09}. This property makes carbon nanomaterials 
promising for transport of spin-polarized currents and for spin-based 
quantum information processing.    

This review provides a brief introduction into the current state of the field of
magnetic materials and nanostructures based on graphene. The possible scenarios 
for the onset of magnetism in graphene nanostructures are illustrated by means of 
a simple theoretical model based on the mean-field Hubbard Hamiltonian. The next 
section briefly reviews the landmark experimental reports in the field. Then 
follows the introduction of the theoretical model and its specific consequences for 
describing the electronic structure and magnetic properties of graphene-based materials. 
The main part of this article applies the theoretical model described and reviews 
both theoretical and experimental published works on magnetic graphene systems 
classified according to their dimensionality: (1) finite graphene nanofragments, 
(2) one-dimensional graphene edges and graphene nanoribbons, and (3) two-dimensional 
graphene and graphite with magnetism induced by the presence of point defects. 
Of these three classes, magnetic graphene edges and nanoribbons will be covered
in more detail since these one-dimensional objects continue to receive special attention 
in the scientific community. Future perspectives of the field are outlined in the 
last section. 

\section{Brief overview of experimental progress}

\begin{figure}[b]
\includegraphics[width=8.5cm]{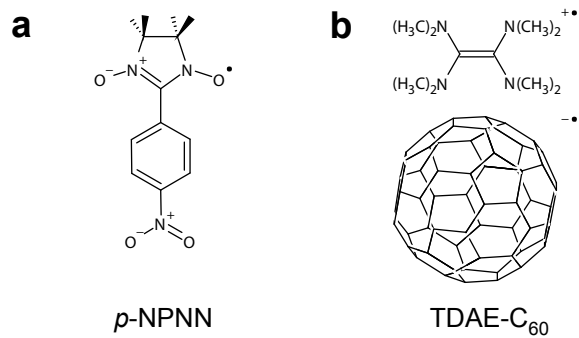}
\caption{\label{tdae} 
Chemical structures of (a) $p$-NPNN and (b) the TDAE-C$_{60}$ charge 
transfer complex. Molecular crystals of these organic compounds
exhibit ferromagnetic ordering with Curie temperatures of 0.6~K and 
16~K, respectively. Each molecular radical unit bears a single 
uncompensated electron spin. In chemical notation the presence of 
aunpaired electron is usually denoted by a thick dot.
}
\end{figure}

The first reproducible experimental reports of magnetism in $p$-block compounds 
were published in 1991 when magnetic ordering was observed in crystalline 
$p$-nitrophenyl nitronyl nitroxide ($p$-NPNN) \cite{Takahashi91,Tamura91} and in 
a charge transfer complex of C$_{60}$ and tetrakis(dimethylamino)ethylene (TDAE) 
\cite{Allemand91}. Molecular structures of these organic materials made of light 
elements only (C, H, N and O) are shown in Figure~\ref{tdae}. In their molecular 
crystals the uncompensated electron spins are localized on weakly 
coupled molecular units. Because of the weak coupling between electron spins, 
the long-range magnetic order is realized only at low temperatures. The two organic
materials mentioned above, $p$-NPNN and TDAE-C$_{60}$, are characterized
by Curie temperatures of 0.6~K and 16~K, respectively. Since 1991 a large number 
of other organic magnetic materials have been examined. In all cases the temperatures 
below which long-range magnetic order is established (Curie temperatures, $T_{\rm C}$,
and N\'eel temperatures, $T_{\rm N}$, in the case of ferromagnetic and antiferromagnetic 
orderings, respectively) were much lower than room temperature, which renders such 
materials useless for practical applications.

The next milestone experiment was reported ten years later when ferromagnetism with 
$T_{\rm C} \approx 500$~K was observed in rhombohedral C$_{60}$ under high pressure
\cite{Makarova01}. This observation, however, demonstrates very well the
controversial character of the field. Five years later several authors retracted 
the original publication since the measured content of magnetic impurities 
was shown to be close to the amount needed to explain the observed
magnetization of the samples \cite{Makarova06b}. In addition, the measured 
$T_{\rm C}$ was found to be very similar to the one of cementite Fe$_3$C. The question of possible 
high-temperature magnetic ordering in C$_{60}$-based materials remains open.
 
\begin{figure}[b]
\includegraphics[width=8.5cm]{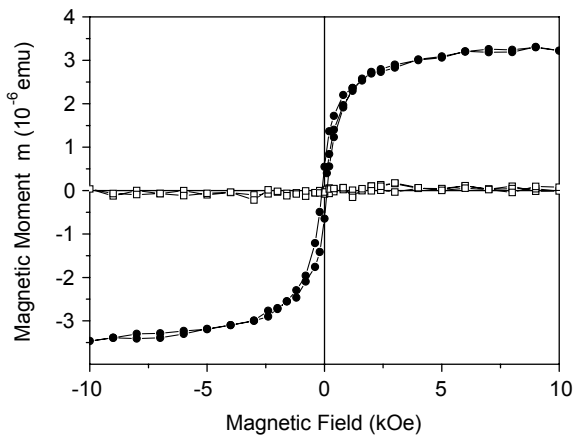}
\caption{\label{esquinazi} 
Magnetic moment of proton irradiated (filled circles) and untreated (empty squares) 
graphite as a function applied magnetic field as measured by Esquinazi {\it et al.}
at $T = 300$~K. The weak hysteresis loop can be recognized. Reprinted 
from \cite{Esquinazi03}. Copyright 2003 by the American Physical Society.
 }
\end{figure}
 
Two years later room-temperature ferromagnetism was observed in highly oriented 
pyrolytic graphite (HOPG) irradiated with high-energy (2.25~MeV) protons \cite{Esquinazi03}.
Figure~\ref{esquinazi} shows the magnetization loop for a proton-irradiated 
sample compared to untreated HOPG. 
Further experimental investigations revealed that the magnetic order in 
proton-bombarded graphite has two-dimensional, that is, graphene-like character \cite{Barzola-Quiquia07} 
and originates from the carbon $\pi$-electron system rather than from possible
$d$-element impurities \cite{Ohldag07}. Interestingly, it was shown that the 
chemical nature of the high-energy particles plays a crucial role in producing 
magnetic ordering. While proton irradiation leads to the onset of ferromagnetism
in irradiated samples, both helium \cite{Esquinazi03} and iron \cite{Barzola-Quiquia08} ions
show no clear effect. On the other hand, the implantation of carbon ions 
was also reported to induce ferromagnetism in HOPG \cite{Xia08}.

A number of reports have also pointed out that even untreated graphite exhibit
ferromagnetism \cite{Kopelevich00,Esquinazi02}. Very recently, by using 
a combination of scanning probe techniques and magnetization measurements, 
\v{C}ervenka and co-authors have shown that the intrinsic ferromagnetism of graphite 
is related to the presence of grain boundaries which can be considered as  
2D periodic networks of point defects \cite{Cervenka09}. Room-temperature
magnetic hysteresis has also been reported for graphene samples produced in bulk 
quantities from graphite using the chemical approaches \cite{Wang09b,Matte09}.

\section{Basic computational approaches}

\subsection{Model Hamiltonians}

The vast majority of computational studies of magnetic carbon nanostructures are 
currently performed using first-principles electronic structure methods based on 
density functional theory (DFT). This approach is now implemented in a large number
of public computer codes and well described in a variety of 
graduate-level textbooks (for instance, see \cite{Koch02,Martin04,Marx09}).
For pedagogical purposes, a simpler approach based on model Hamiltonians will be 
adopted in this review article. Moreover, it will be demonstrated below that 
such simplified models very often allow deeper understanding of the results obtained. 

A simple model which is widely used for studying magnetic effects in $sp^2$ carbon 
materials is the one-orbital mean-field Hubbard model. This model considers 
only the $\pi$-symmetry electronic states which are formed by the unhybridized 
$p_z$ atomic orbitals of $sp^2$ carbon atoms. As shown in Figure~\ref{fig1}(b) 
the low-energy electronic states have $\pi$-symmetry and thus play the dominant role 
in the properties of graphene systems. The Hubbard model Hamiltonian can be partitioned
into two parts,
\begin{equation}
\label{eq1}
 {\mathcal H} = {\mathcal H}_0 + {\mathcal H}'.
\end{equation}
The first term is the nearest-neighbor tight-binding Hamiltonian
\begin{equation}
\label{eq2}
	{\mathcal H}_0 = - t \sum_{\la i,j \ra, \sigma} [ c_{i\sigma}^\dagger c_{j\sigma} + {\rm h.c.} ],
\end{equation}
in which the operators $c_{i\sigma}$ and $c_{i\sigma}^\dagger$ annihilate and create an electron with spin $\sigma$ at site $i$, 
respectively. The notation $\la \cdot,\cdot \ra$ stands for the pairs of nearest-neighbor atoms; `h.c.' 
is the Hermitian conjugate counterpart. The well established hopping integral $t \approx 2.7$~eV 
defines the energy scale of the Hamiltonian. This physical model is equivalent to 
the H\"uckel method familiar to chemists. From the computational point of view, the 
Hamiltonian matrix is determined solely by the atomic structure: the off-diagonal 
matrix elements $(i,j)$ and $(j,i)$ are set to $-t$ when atoms $i$ and $j$ are covalently bonded, 
and to 0 otherwise. In a neutral graphene system each $sp^2$ carbon atom contributes one 
$p_z$ orbital and one $\pi$ electron. The $\pi$-electron system is thus called half-filled. 
The spectrum of the eigenvalues of tight-binding Hamiltonian matrix exhibit electron-hole symmetry, 
{\it i.e.} it is symmetric with respect to zero energy. In other words, in a neutral graphene system for each eigenvalue 
$\e < 0$ corresponding to an occupied (bonding) state, there is an unoccupied (anti-bonding) 
state with $\e^\star = -\e$. The states with $\e = 0$ are called zero-energy states 
(also referred to as non-bonding or midgap states). 

The nearest-neighbor tight-binding model has proved to describe accurately the electronic 
structure of graphene, carbon nanotubes and other non-magnetic $sp^2$ carbon materials. 
However, electron-electron interactions have to be introduced in some form in order to 
describe the onset of magnetism. Within the Hubbard model these interactions are 
introduced through the repulsive on-site Coulomb interaction
\begin{equation}
\label{eq3}
 {\mathcal H}' = U \sum_i n_{i\ua} n_{i\da},
\end{equation}
where $n_{i\sigma}=c_{i\sigma}^\dagger c_{i\sigma}$ is the spin-resolved electron density at site $i$; the parameter $U>0$  defines the 
magnitude of the on-site Coulomb repulsion. This model considers only the short-range Coulomb 
repulsion, that is, two electrons interact only if they occupy the $p_z$ atomic orbital of 
the same atom. Despite its apparent simplicity, this term is no longer trivial from the 
computational point of view. The mean-field approximation
\begin{equation}
\label{eq4}
 {\mathcal H}'_{\rm mf} =  U \sum_i \left( n_{i\ua} \la n_{i\da} \ra  + \la n_{i\ua} \ra n_{i\da} - \la n_{i\ua} \ra \la n_{i\da} \ra \right), 
\end{equation}
allows us to overcome this difficulty. Here, a spin-up electron at site $i$ interacts with the average spin-down electron 
population $\la n_{i\da} \ra$ at the same site, and vice versa. This 
mean-field model represents a variation of the unrestricted Hartree-Fock method \cite{Szabo82}. 
From the computational point of view, the electron-electron interaction term affects only the diagonal 
elements of the Hamiltonian matrix. The diagonal elements of the spin-up and spin-down 
blocks now depend on the unknown $\la n_{i\da} \ra$ and 
$\la n_{i\ua} \ra$, respectively. The problem can be solved 
self-consistently starting from some initial values of $\la n_{i\sigma} \ra$ which
can be chosen randomly. However, one has to keep in mind that in certain cases
the broken-symmetry (antiferromagnetic) solutions can be obtained only if the 
initial guess breaks the spin-spatial symmetry \cite{Yazyev08d}. The process of (1) calculation of 
the matrix elements of the Hamiltonian matrix, (2) its diagonalization and (3) the computation 
of updated spin densities is then repeated iteratively until all values of 
$\la n_{i\sigma} \ra$ are converged. The final self-consistent solution provides the 
spin densities
\begin{equation}
\label{eq5}
 M_i = \frac{\la n_{i\ua} \ra - \la n_{i\da} \ra}{2},
\end{equation}
at each atom $i$ and the total spin of the system  $S = \sum_i M_i$. For a given graphene 
structure both local and total spins (magnetic moments) depend exclusively on the dimensionless 
parameter $U/t$. 

After the model has been introduced, the following three critical questions can be asked. 
(1) Is the one-orbital approximation accurate enough compared to the methods considering
all electrons? (2) Which value of the empirical parameter $U/t$ should be used? (3) Is the mean-field
approximation justified for graphene based systems? 

(1) It has been shown that the results of mean-field Hubbard model calculations 
correspond closely to the ones obtained using first-principles methods if 
the parameter $U/t$ is chosen appropriately \cite{Fernandez-Rossier07,Pisani07,Gunlycke07}. 
The first-principles methods either treat all electrons on equal footing or disregard 
the localized atomic core states which are not important in most cases. One notable exception
is the calculation of hyperfine interactions. In this case the spin polarization 
of the $1s$ atomic core states of carbon atoms has significant contribution to the Fermi 
contact hyperfine couplings \cite{Yazyev05,Yazyev08b}. 
Otherwise, the results of DFT calculations performed using a generalized-gradient-approximation 
family exchange-correlation functional are best reproduced when $U/t \approx 1.3$.
The results of the local-spin-density approximation calculations are best fitted using $U/t \approx 0.9$ \cite{Pisani07}. 

(2) Ideally, the empirical parameter $U/t$ must be estimated using experimental knowledge.
Unfortunately, there are no direct experiments performed on magnetic graphene systems which 
would allow to estimate $U/t$. Magnetic resonance studies of neutral soliton states in 
{\it trans}-polyacetylene, a one-dimensional $sp^2$ carbon polymer which can be viewed
as a minimum-width zigzag graphene nanoribbon, give the range of values $U \sim 3.0$$-$3.5~eV   
\cite{Thomann85,Kuroda87}. This interval corresponds to $U/t \sim 1.1$$-$1.3 which also 
makes us confident in the results of the generalized-gradient-approximation DFT calculations. Increasing $U/t$ 
leads to the enhancement of magnetic moments. The range of meaningful magnitudes is 
limited by $U/t \approx 2.23$ above which the ideal graphene undergoes a Mott-Hubbard
transition into an antiferromagnetically ordered insulating state \cite{Sorella92}. 
In the computational examples considered below a value of $U/t = 1.2$ will be used.

(3) This question is the most difficult to answer. A comparison of the mean-field results
with the ones obtained using exact diagonalization and quantum Monte Carlo simulations
illustrates the validity of this approximation for the relevant values of $U/t$ \cite{Heldner09}. 
Magnetic graphene materials and nanostructures need not be considered as strongly 
correlated systems.

\subsection{Counting rules}

There are two important consequences coming from the model Hamiltonians we have introduced. The honeycomb lattice of graphene is a 
{\it bipartite} lattice. That is, it can be partitioned into two mutually interconnected sublattices 
$A$ and $B$ (see Fig.~\ref{fig1}(a)). Each atom belonging to sublattice $A$ is connected 
to the atoms in sublattice $B$ only, and vice versa. Moreover, the graphene systems whose 
faces are hexagons are called benzenoid (or honeycomb) systems. 
Carbon atoms in such systems have either three or two nearest neighbors. The class of 
benzenoid systems is a subclass of bipartite systems.

The spectrum of the tight-binding Hamiltonian of a honeycomb system can be analyzed using 
a mathematically rigorous approach of the benzenoid graph theory \cite{Fajtlowicz05}. 
An important result for us is that this theory is able to predict the number of zero-energy 
states of the nearest-neighbor tight-binding Hamiltonian in a `counting rule' fashion. The number of such states is equal to 
the graph's nullity
\begin{equation}
\label{eq6}
  \eta = 2\alpha -N,
\end{equation}
where $N$ is the total number of sites and $\alpha$ is the maximum possible number of non-adjacent sites,
{\it i.e.} the sites which are not the nearest neighbors to each other. 
The onset of magnetism in the system is determined by the so-called Stoner criterion which
refers to the competition of the 
exchange energy gain and the kinetic-energy penalty associated with the spin-polarization 
of the system \cite{Mohn03}. The gain in exchange energy is due to the exchange splitting 
of the electronic states subjected to spin-polarization \cite{Palacios08}
\begin{equation}
\label{eq7}
  \Delta_S = \e_\ua - \e_\da = \frac{U}{2}\sum_i n_i^2,
\end{equation}
where  $\sum_i n_i^2$ is the inverse participation ratio, a measure of the degree of localization of the 
corresponding electronic state. The kinetic-energy penalty is proportional to the energy of 
this state. Thus, the zero-energy states undergo spin-polarization at any $U > 0$ irrespective 
of their degree of localization. One can view spin-polarization as one of the mechanisms 
for escaping an instability associated with the presence of low-energy electrons in the 
system. Other mechanisms, such as the Peierls distortion, were shown to be inefficient in the case of
graphene nanostructures \cite{Pisani07}.
	
	Although the benzenoid graph theory is able to predict the occurrence of zero-energy states,
it is not clear how the electron spins align in these states. The complementary knowledge 
is supplied by Lieb's theorem \cite{Lieb89} which determines the total spin of a bipartite system 
described by the Hubbard model. This theorem states that in the case of repulsive 
electron-electron interactions ($U > 0$), a bipartite system at half-filling has the ground 
state characterized by the total spin
\begin{equation}
\label{eq8}
  S = \frac{1}{2} |N_A-N_B|,
\end{equation}
where $N_A$  and $N_B$  are the numbers of sites in sublattices $A$ and $B$, respectively. The ground state 
is unique and the theorem holds in all dimensions without the necessity of a periodic lattice 
structure. Importantly, the two counting rules are linked by the following relation, $\eta \geq |N_A-N_B|$.
	
In the following section, the application of these two simple counting rules will be 
demonstrated on small graphene fragments and compared to the results of numerical 
calculations.

\section{Finite graphene fragments - a simple illustration}

\begin{figure}[h]
\includegraphics[width=8.5cm]{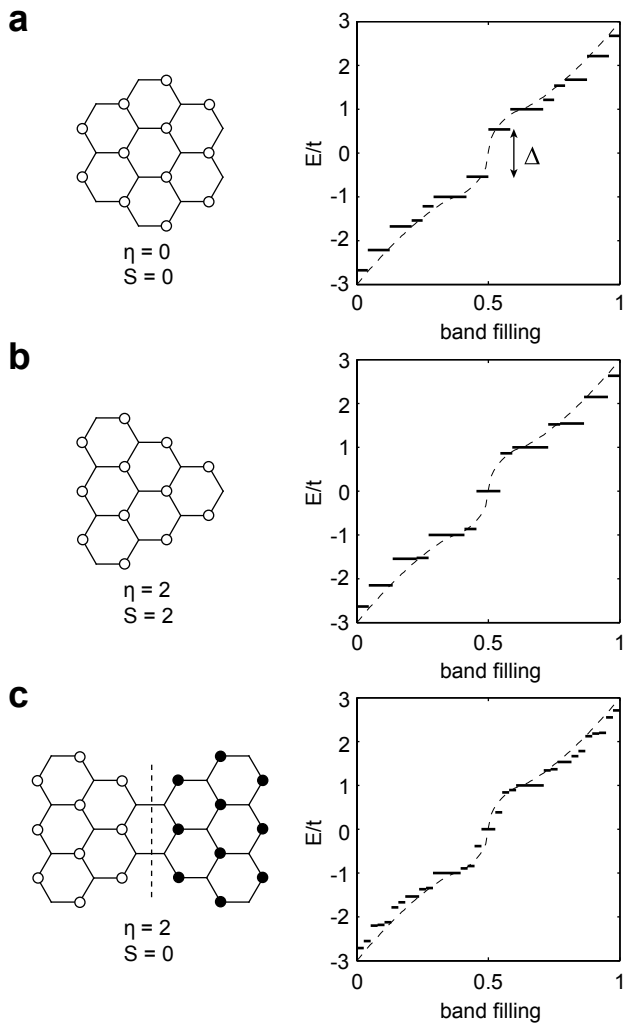}
\caption{\label{fig2} 
Atomic structures and tight-binding energy spectra of three graphene fragments: 
(a) coronene, (b) triangulane, and (c) a bowtie-shaped fragment (``Clar's goblet''). 
Non-adjacent sites are labeled by circles. Empty and filled circles correspond
to sublattice $A$ and sublattice $B$, correspondingly. Tight-binding energies 
are plotted as a function of band filling. Dashed line corresponds to the energy spectrum 
of ideal graphene.
}
\end{figure}

Let us first try to understand the origin of magnetism in finite graphene 
fragments (also referred to as nanoflakes, nanoislands or nanodisks) as
a function of their shape and size. Three simple examples of nanometer sized
graphene fragments are shown in Figure~\ref{fig2}. From the point of view of 
single-orbital physical models, only the connectivity of the $\pi$-electron 
conjugation network is important. Such $\pi$-systems may constitute only small 
parts of more complex molecules or bulk materials. In simplest case, the
$\pi$-electron networks shown in Figure~\ref{fig2} can be realized in the 
corresponding all-benzenoid polycyclic aromatic hydrocarbon (PAH) molecules 
with the edges of the fragments being passivated by hydrogen atoms. Each carbon atom
at the edge of the fragment is bonded to one hydrogen atom such that
all carbon atoms are $sp^2$-hybridized. Current progress
in synthesizing such molecules and understanding their properties has
recently been reviewed \cite{Wu07}.

The hexagonal graphene fragment shown in Figure~\ref{fig2}(a) is thus equivalent 
to the {\it coronene} molecule. For this fragment, the number of sites belonging 
to the two sublattices is equal, $N_A=N_B=12$. The number of non-adjacent 
sites is maximized when all atoms belonging to either of the two sublattices are 
selected, {\it i.e.} $\alpha = 12$. Thus, both the number of zero-energy states 
$\eta$ and the total spin $S$ are zero. The tight-binding model predicts 
a wide band gap of $1.08t\approx 3.0$~eV for this graphene molecule. 
As expected, the mean-field Hubbard model solution for this fragment 
does not reveal any magnetism.

\begin{figure}[b]
\includegraphics[width=8.5cm]{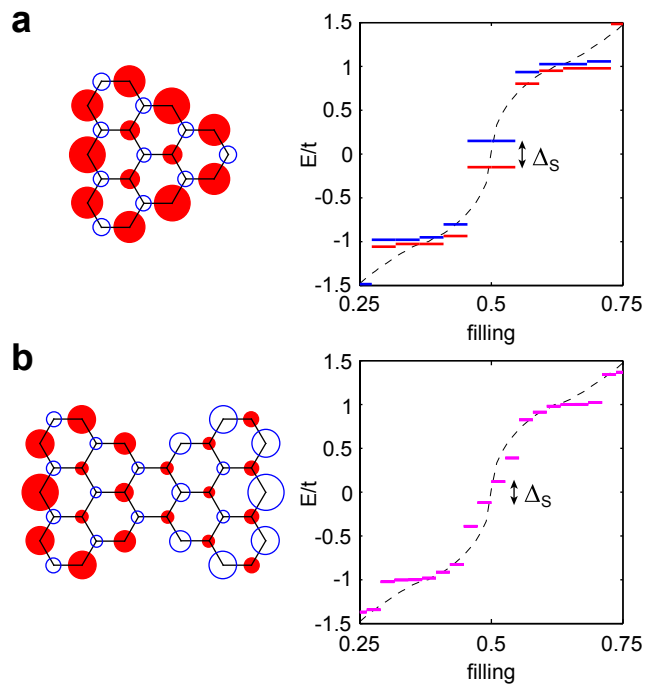}
\caption{\label{fig3}  
Local magnetic moments and spin-resolved energy levels obtained through the 
mean-field Hubbard model calculations for (a) triangulane and (b) the bowtie-shaped 
graphene fragment ($U/t = 1.2$). Area of each circle is proportional to the magnitude 
of the local magnetic moment at each atom. Filled (red) and empty (blue) circles 
correspond to spin-up and spin-down densities. Energy levels energies 
are plotted as a function of band filling. Dashed line corresponds to the energy spectrum 
of ideal graphene.Red and blue levels correspond spin-up and spin-down channels, 
respectively. In the case of bowtie fragment the energies in the two channels are 
identical (shown in magenta). 
}
\end{figure}

The second graphene fragment shown in Figure~\ref{fig2}(b) has triangular shape. 
It is not surprising that the corresponding hypothetical PAH molecule is called 
{\it triangulane}. Unlike coronene, the two sublattices of this triangular
fragment are no longer equivalent:  $N_A=12$ and $N_B=10$. 
The unique choice maximizing the number of non-adjacent sites is achieved by 
selecting the atoms belonging to the dominant sublattice $A$, {\it i.e.} $\alpha=N_A=12$. 
Thus, the benzenoid graph theory predicts the presence of two zero-energy states 
on sublattice $A$. Lieb's theorem predicts the $S = 1$ (spin-triplet) ground state or, 
equivalently, a magnetic moment of 2~$\mu_{\rm B}$ per molecule. 
The two low-energy electrons populate a pair of zero-energy states according to 
Hund's rule, that is, their spins are oriented parallel to each other. The mean-field Hubbard model 
results for this system at half-filling are shown in Figure~\ref{fig3}(a). 
One can see that spin-polarization lifts the degeneracy of the zero-energy 
electronic states and opens an energy gap $\Delta_S = 0.30t\approx 0.8$~eV. 
The system is stabilized by spin-polarization. Most of the spin-up 
electron density localized on the atoms in sublattice $A$ (see Fig.~\ref{fig3}(a)) originates from the two 
electrons populating the non-bonding states. However, one can notice an appreciable 
amount of spin-down density on the atoms in sublattice $B$ which is compensated by an 
equivalent contribution of the spin-up density in sublattice $A$. The occurrence of 
the induced magnetic moments is a manifestation of the spin-polarization effect which is
related to the exchange interaction of the fully populated states with the two unpaired 
electrons.

The third bowtie-shaped graphene molecule shown in Figure~\ref{fig2}(c) is composed 
of two triangulane fragments sharing one hexagon.  For this system Lieb's theorem 
predicts the spin-singlet ground state ($N_A=N_B=19$). However, the choice of 
the set of atoms which maximizes the number non-adjacent sites is less evident in 
this case. Figure~\ref{fig2}(c) shows such a selection ($\alpha=20$) which
involves the atoms belonging to both sublattice $A$ and sublattice $B$ in the left 
and right parts of the structure. These atoms are marked differently in the figure.  
Hence, there are $\eta=2\times20-38=2$ zero-energy states as confirmed by the 
tight-binding calculation. The zero-energy states are spatially segregated in the 
two triangular parts of the molecule \cite{Wang09}. To satisfy the spin-singlet 
ground state, the two zero-energy states have to be populated by two electrons with 
oppositely oriented spins. In other words, the ground electronic configuration 
breaks spin-spatial symmetry and exhibits antiferromagnetic ordering. This 
result can be verified by mean-field Hubbard model calculations as shown in 
Figure~\ref{fig3}(b). It can be argued that this example violates Hund's rule. 
However, one has to keep in mind that each of the two non-bonding states is 
localized within one of the graphene sublattices. That is, there are two electronic 
sub-bands, each populated by electrons according to Hund's rule. The coupling 
between the electron spins in these two sub-bands is antiferromagnetic due to 
the superexchange mechanism \cite{Kramers34,Anderson50}.

\begin{figure}[b]
\includegraphics[width=8.5cm]{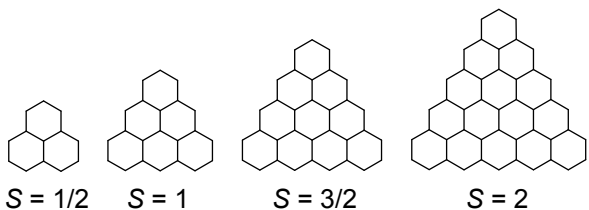}
\caption{\label{scaling} 
Evolution of the total spin of triangular graphene fragments with size.
}
\end{figure}

The two counting rules can be applied to larger graphene fragments. It was shown 
that the total spin of triangular fragments with edges cut along the zigzag
direction scales linearly with fragment size \cite{Wang08,Fernandez-Rossier07,Ezawa07} 
as illustrated in Figure~\ref{scaling}. The average magnetic moment per carbon atom 
thus decays with increasing the system size. The evolution of magnetic properties 
with increasing size for hexagonal fragments with edges cut along the same zigzag 
direction is less trivial. It has been shown theoretically that above some critical 
size the system undergoes a transition into a broken-symmetry antiferromagnetic 
state \cite{Fernandez-Rossier07}. The critical size itself depends strongly on the value 
of $U/t$. However, it is easier to explain the origin of this behavior in large systems 
from the standpoint of edge magnetism, which will be explained in the next section.       

\begin{figure}
\includegraphics[width=8.5cm]{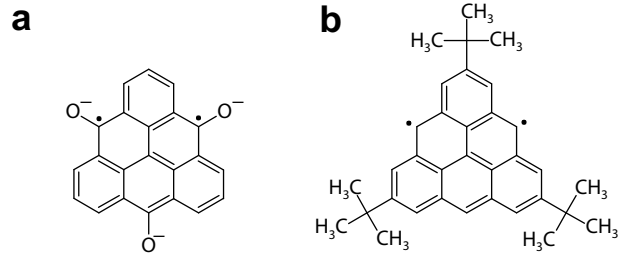}
\caption{\label{triangulanes} 
Chemical derivatives of triangulane synthesized by \cite{Allinson95} and \cite{Inoue01}.
Their spin-triplet ground state has been verified by means of electron spin resonance measurements.}
\end{figure}

Finally, a few words have to be said about the possibility of realizing in practice 
the magnetic graphene fragments we have discussed. It is expected that such magnetic
systems are more reactive than the non-magnetic polyaromatic molecules. Although 
triangulane itself has never been isolated, successful synthesis of its chemical 
derivatives shown in Figure~\ref{triangulanes} has been reported \cite{Inoue01,Allinson95}. 
The spin-triplet ground state of these chemical compounds was verified by the
electron spin resonance measurements. In principle, this example can be considered as an indirect
proof of edge magnetism in graphene systems, at least in finite fragments
produced by means of the chemical bottom-up approach. The synthesized triangulane derivatives
are reactive molecules, but nevertheless can be handled in common organic solvents and stored
for many months at room temperature provided the solution is isolated from atmospheric oxygen
\cite{Allinson95}. Larger magnetic triangular molecules have not been synthesized
so far. The PAH molecule corresponding to the considered bowtie fragment
was hypothesized by Eric Clar and named ``Clar's goblet'' after him \cite{Clar72}.
Attempts to synthesize this molecule have failed \cite{Clar72b}.

The examples shown above illustrate how three different magnetic scenarios
can be realized in very simple finite graphene systems. These examples also
provide a way for designing nanostructures with predefined magnetic interactions,
a highly useful tool for developing novel spintronic devices. The value of this approach has already been demonstrated 
by the proposal of reconfigurable spintronic logic gates exploiting the strong 
antiferromagnetic couplings in the bowtie-shaped graphene fragments \cite{Wang09}. Several 
devices for controlling spin-currents based on triagular graphene fragments have also 
been described recently \cite{Ezawa09}. A number of other intriguing properties
predicted for zero-dimensional graphene fragments are potentially interesting 
from both fundamental and applied points of view \cite{Akola08,Ezawa08,Ezawa09b}.  

\section{Zigzag edges and nanoribbons}

\subsection{Physical mechanism of edge magnetism}

\begin{figure}
\includegraphics[width=8.5cm]{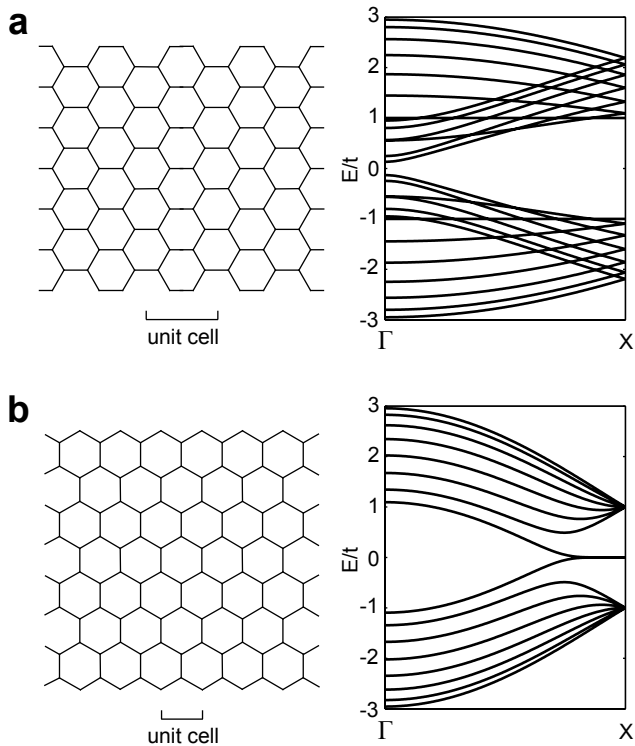}
\caption{\label{fig4} 
Atomic structures and tight-binding band structures of (a) armchair and (b) 
zigzag graphene nanoribbons. Unit cells of the graphene nanoribbons are shown.
}
\end{figure}

As one moves on towards larger graphene fragments or infinite systems, the application 
of counting rules becomes impractical. An alternative approach considers the effects of 
edges of graphene nanostructures which can be conveniently modeled using one-dimensional 
periodic strips of graphene. Such models are commonly referred to as {\it graphene 
nanoribbons}. There are two high-symmetry crystallographic directions 
in graphene, {\it armchair} and {\it zigzag}, as shown in Figure~\ref{fig1}(a). Cutting graphene 
nanoribbons along these directions produces armchair and zigzag nanoribbons, respectively  (Fig.~\ref{fig4}).

The band structures of armchair and zigzag nanoribbons are remarkably
different. Figure~\ref{fig4}(a) shows the tight-binding band structure of a 
$\sim$1.5~nm wide armchair nanoribbon. For this particular armchair graphene
nanoribbon, introducing a pair of parallel armchair edges opens a gap of 
$0.26t$. The nearest-neighbor tight-binding model predicts either metallic 
or semiconducting behavior for armchair nanoribbons \cite{Ezawa06,Son06a,Nakada96,Barone06,Brey06,Peres06}, 
and the two situations alternate as the nanoribbon's width increases. The band gap
of semiconducting graphene nanoribbons decreases with increasing 
width. In the case of metallic nanoribbons two bands cross the Fermi level at the  
$\Gamma$ point. No magnetic ordering is predicted in this case.

Within the same model all zigzag graphene nanoribbons are metallic and feature a 
flat band extending over one-third of the one-dimensional Brillouin zone at 
$k \in \left({2\pi}/{3a};{\pi}/{a}\right)$ ($a = 0.25$~nm is the unit cell 
of the zigzag edge) as shown in Figure~\ref{fig4}(b). Strictly speaking, 
the flat band does not correspond to zero-energy states, but rather to the states 
with energies the approach zero with increasing 
nanoribbon width. The low-energy states are localized at the edge and decay quickly 
in the bulk.

\begin{figure}[b]
\includegraphics[width=8.5cm]{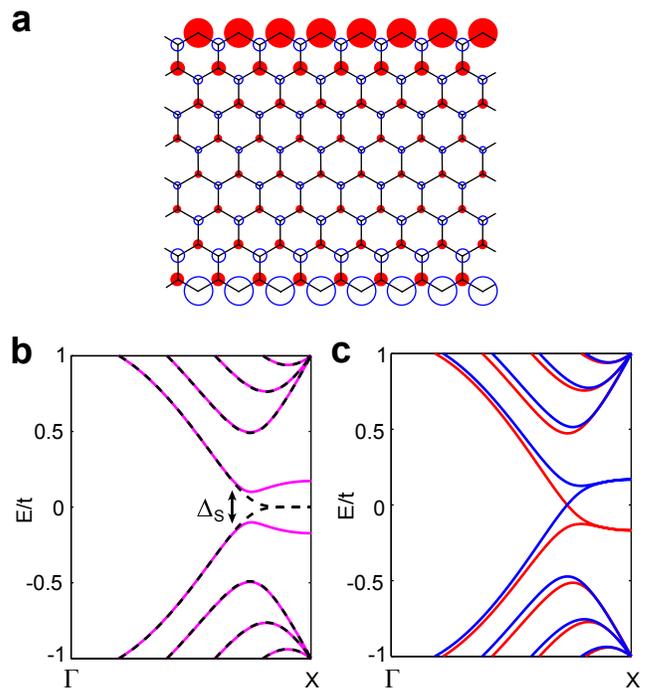}
\caption{\label{fig5} 
(a) Local magnetic moments in a zigzag graphene nanoribbon calculated using 
the mean-field Hubbard model ($U/t = 1.2$). Area of each circle 
is proportional to the magnitude of the local magnetic moment at each atom. 
Filled (red) and empty (blue) circles correspond to spin-up and spin-down 
densities, respectively. (b) Mean-field Hubbard-model band structure 
(solid magenta lines) compared to the tight-binding band structure (dashed lines) for
the solution shown in panel (a). The band structures for spin-up and spin-down 
electrons are equivalent. (c) Mean-field Hubbard-model band structure
for the same graphene nanoribbon with the ferromagnetic mutual orientation of
the edge spins.  The band structures for the majority-spin electrons and
the minority-spin electrons are shown as red and blue lines, respectively. 
}
\end{figure}

High density of low-energy electronic states suggests a possibility 
of magnetic ordering. Indeed, the mean-field Hubbard model solution for this 
system reveals magnetic moments localized at the edges as shown in Figure~\ref{fig5}(a). The localized 
magnetic moments display ferromagnetic ordering along the zigzag edge while the mutual 
orientation of the magnetic moments localized at the opposite edges is antiparallel \cite{Fujita96,Son06b,Fernandez-Rossier08}.
Thus, the net magnetic moment of a zigzag nanoribbon is zero in agreement
with Lieb's theorem ($N_A = N_B$). The band structure corresponding to the mean-field Hubbard model 
solution is compared to the tight-binding band structure in Figure~\ref{fig5}(b). 
The introduced electron-electron interactions open a band gap across the whole flat-band segment
turning the system into a semiconductor ($\Delta_S = 0.20t$ at $U/t = 1.2$).
The spin-polarization almost does not affect the electronic states at higher energies.
The band structures for the two spin channels are equivalent, but spin-spatial 
symmetry is broken.  

\begin{figure}[b]
\includegraphics[width=8.5cm]{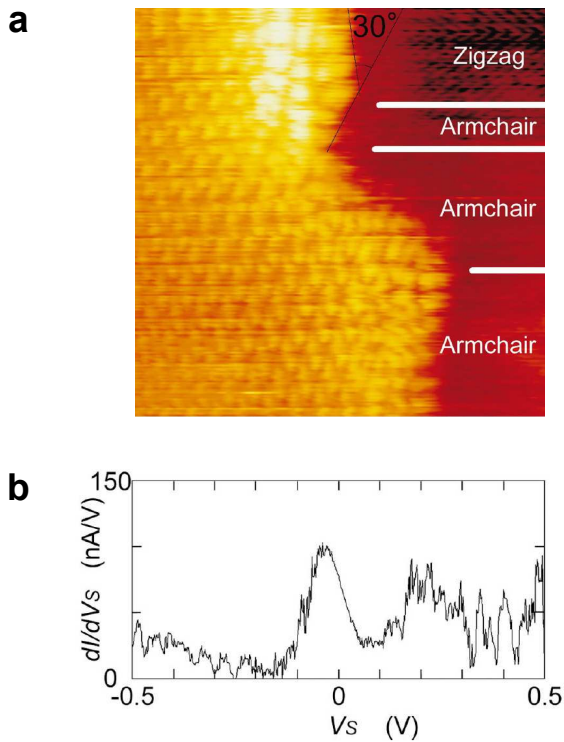}
\caption{\label{enoki} 
(a) An atomically resolved STM image showing the presence of 
both armchair and zigzag graphene edges. (b) Typical STS curve 
measured at a zigzag edge reveals the peak related to zero-energy 
states. Reproduced from \cite{Kobayashi05}.
Copyright 2005 by the American Physical Society.
 }
\end{figure}

The coupling between the magnetic edges can be ascribed to the superexchange mechanism as in the case  
of bowtie graphene fragment considered in the previous section. The magnitude of 
the antiferromagnetic coupling shows a $w^{-2}$ dependence as a function 
of nanoribbon width $w$ \cite{Jung09a}. The interedge magnetic coupling strength 
of $\sim$25~meV has been calculated from first principles for a $\sim$1.5~nm
wide nanoribbon \cite{Pisani07}. Unlike the antiferromagnetic ground state,
a zigzag graphene nanoribbon with ferromagnetic interedge orientation is
a metal with two bands crossing the Fermi level at $k \approx {2\pi}/{3a}$ (Fig.~\ref{fig5}(c)).
The possibility of switching between the two states was exploited in a proposal
of a graphene-based magnetic sensor \cite{Munoz-Rojas09} described in the next 
section. The coupling between the magnetic moments localized at the edges 
can be controlled by means of either electron or hole doping of the nanoribbons 
\cite{Sawada09,Jung09b}. High doping levels eventually suppress magnetism
since the flat band shifts away from the Fermi level, thus eliminating the 
electronic instability associated with the presence of low-energy electrons \cite{Jung09b}. 

It is worth mentioning that at the time this review was written, no direct proof of 
edge magnetism in graphene has been reported. However, the presence of localized 
low-energy states at zigzag edges of graphene has been verified by means of scanning 
tunneling microscopy \cite{Kobayashi05,Kobayashi06}.

\subsection{Possible applications in spintronics}

\begin{figure}[b]
\includegraphics[width=8.5cm]{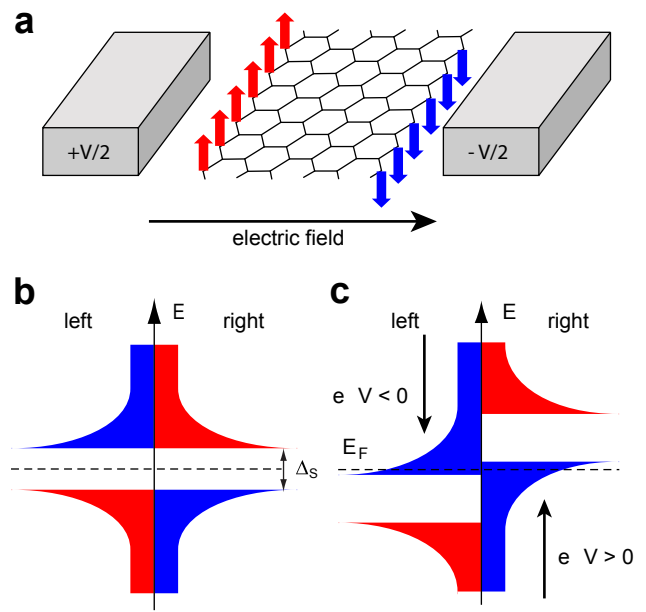}
\caption{\label{fig6} 
Scheme of electric-field-induced half-metallicity in zigzag graphene nanoribbons. 
(a) Electric field is applied across the nanoribbon, from left edge (spin-up, 
red arrows) to right edge (spin-down, blue arrows). (b) Schematic representation 
of the spin-resolved local density of states for the opposite edges at zero applied 
field. (c) Applied electric field closes the band gap at the Fermi level $E_F$ 
for spin-down electrons selectively.}
\end{figure}

It has been realized that the intriguing magnetic properties of graphene nanostructures 
may find applications in spintronics. The pioneering idea was introduced 
by Son, Cohen and Louie, who have predicted that external electric fields induce {\it half-metallicity} 
in zigzag graphene nanoribbons \cite{Son06b}. The half-metallicity refers to the coexistence 
of a metallic state for electrons with one spin orientation and an insulating state for electrons 
with the opposite spin orientation. An electric field is applied across the nanoribbon 
as shown in Figure~\ref{fig6}(a). At zero field the system is characterized by the energy gap $\Delta_S$ 
for the spin-polarized states localized at both edges (Fig.~\ref{fig6}(b)). An applied electric 
field breaks the symmetry and closes the gap for one of the spin directions selectively (Fig.~\ref{fig6}(c)). 
The critical field required for inducing the half-metallicity is $3.0/w$ Volts, where $w$ is the 
nanoribbon width in \AA. The direction of the applied electric field defines the spin 
channel with metallic conductivity. If realized in practice, this simple device 
would offer efficient electrical control of spin transport -- a highly desirable component 
for spintronics.

\begin{figure}
\includegraphics[width=8.5cm]{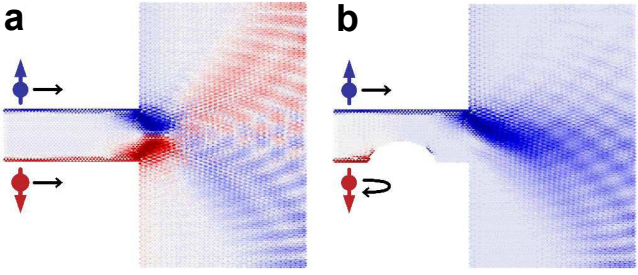}
\caption{\label{wimmer} 
Spin injection from (a) an ideal zigzag graphene nanoribbon and (b) a nanoribbon
with a defective bottom edge into $n$-doped graphene \cite{Wimmer08}.
Nonequilibrium densities for spin-up and spin down-electrons are shown in blue and
red, respectively. Copyright 2008 by the American Physical Society.
}
\end{figure}

In addition to the device described, several other approaches for controlling the spin 
transport in graphene nanostructures have been proposed. One of them exploits 
disorder for achieving the goal; an example from \cite{Wimmer08} is shown in 
Figure~\ref{wimmer}. Electric current flowing along the edges of zigzag 
nanoribbon injects spin-polarized electrons into a graphene reservoir 
(Fig.~\ref{wimmer}(a)). However, the net spin-polarization of the current
is zero due to the antiferromagnetic coupling between the two equivalent 
edges. Then, an extended defect is introduced into one of the edges as shown 
in Figure~\ref{wimmer}(b). The defect both quenches magnetic moments and scatters 
the carriers at the rough edge. However, conduction at the opposite edge
remains unaffected, thus allowing for injecting a current with a net spin 
polarization. Other proposals based on defect and impurity engineering of 
spin transport in graphene nanoribbons have been reported \cite{Cantele09,Lakshmi09,Park09,Rocha09}.

\begin{figure}
\includegraphics[width=8.5cm]{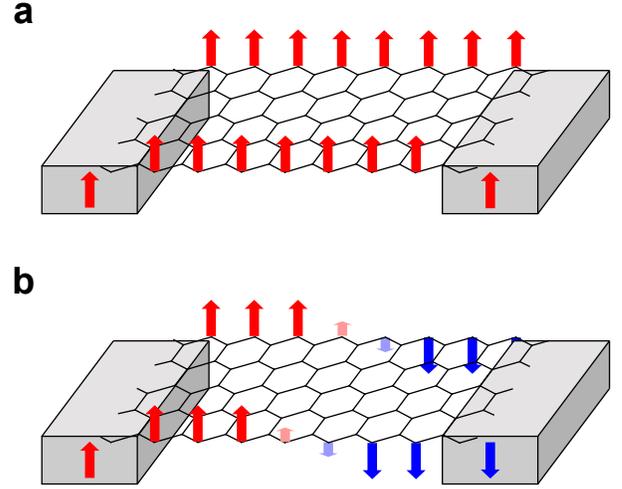}
\caption{\label{kim} 
Scheme of magnetoresistive device based on zigzag graphene nanoribbon connecting
two ferromagnetic contacts \cite{Kim08}. The low-resistance 
configuration (a) and the high-resistance configuration (b) of the device 
correspond to parallel and antiparallel orientations of the magnetic moments
of the two ferromagnetic leads, respectively. Arrows denote the magnetic 
moments of both graphene edges and ferromagnetic leads.
 }
\end{figure}

Magnetic graphene nanostructures were also proposed as components of 
magnetoresistive junctions. Such devices are currently used as 
magnetic field sensors, {\it e.g.} in the read heads of hard disk drives. 
Typical magnetoresistive junctions involve ferromagnetic metal layers 
separated by a nonmagnetic spacer layers, {\it e.g.} Co layers separated
by a non-magnetic Cu layer, or bcc Fe layers separated by a few-nanometers thick layer
of crystalline MgO. A crucial characteristic of such spintronic 
devices is their magnetoresistance ratio (MR) which shows
the change in electric resistance as a function of the relative 
orientation of the magnetization of two ferromagnetic layers \cite{Heiliger06}. This 
quantity can be defined as
\begin{equation}
 {\rm MR} = \frac{R^{\rm AP}-R^{\rm P}}{{\rm min}(R^{\rm P},R^{\rm AP})}\times 100\%,
\end{equation}
where $R^{\rm P}$ and $R^{\rm AP}$ are the resistances for parallel and antiparallel 
relative orientations of the magnetic moments of the layers. Magnetoresistive
devices with high magnitudes of MR are demanded by the future nanoscale 
electronics. It has been predicted that a zigzag graphene nanoribbon placed 
between two ferromagnetic contacts constitutes a magnetoresistive
junction with very high values of magnetoresistance ratio \cite{Kim08}.
The low-resistance state of 
the proposed device corresponds to the parallel configuration in which
the magnetic moments at graphene edges are coupled ferromagnetically
to each other. The ferromagnetic coupling is enforced by the strong 
interaction with the magnetic moments of ferromagnetic contacts (Fig.~\ref{kim}(a)).
In the antiparallel configuration (Fig.~\ref{kim}(b)), the magnetic
graphene nanoribbon develops a domain-wall arrangement of edge spins
with high resistance. It is worth mentioning that spin-transport measurements
in micrometer-scale lateral graphene devices contacted by ferromagnetic 
electrodes have been carried out experimentally \cite{Hill06,Tombros07,Tombros08,Jozsa08}. However, the 
magnetoresistance effect observed in these experiments is due to
the long spin-diffusion lengths in graphene.

\begin{figure}
\includegraphics[width=8.5cm]{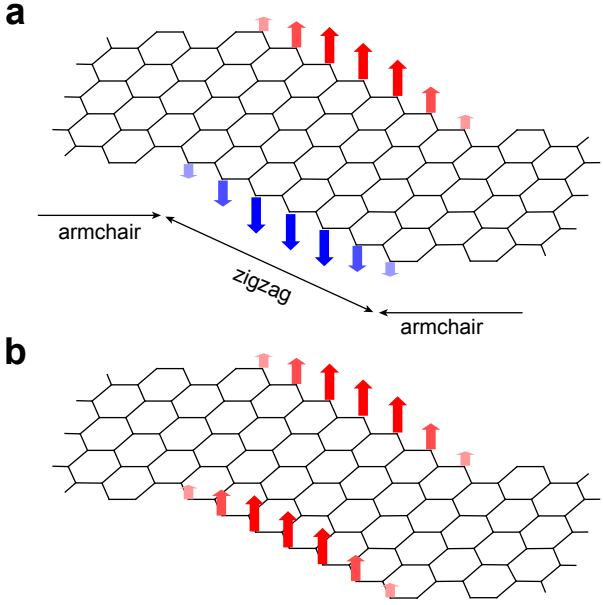}
\caption{\label{federico} 
All-graphene magnetoresistive device proposed in \cite{Munoz-Rojas09}.
The device represents an armchair-zigzag-armchair nanoribbon junction.
The high-resistance antiparallel configuration (a) switches into
a low-resistance parallel configuration (b) in a strong enough applied
magnetic field.     
}
\end{figure}

An all-graphene device based on the armchair-zigzag-armchair nanoribbon
junction has also been predicted to show magnetoresistance effect \cite{Munoz-Rojas09}. 
In zero applied magnetic field the magnetic coupling between the opposite 
zigzag edge segments is antiferromagnetic (Fig.~\ref{federico}(a)) and, hence, 
electric resistance is high due to the gapped electronic state. Sufficiently
strong magnetic fields favor the parallel configuration (Fig.~\ref{federico}(b)) which shows
a lower resistance due to the spin-polarized edge states crossing the Fermi 
level (Fig.~\ref{fig5}(c)). This ultrasmall device thus acts a magnetic field sensor capable to detect magnetic
fields from few hundreds of Gauss to several Tesla at low temperatures. 

Epitaxial graphene and the isostructural hexagonal boron nitride ($h$-BN) have 
been proposed as efficient ultrathin non-magnetic 
spacers for traditional multilayer magnetoresistive junctions \cite{Karpan07,Karpan08,Yazyev09}. 
Unlike in the device shown in Fig.~\ref{kim}, mono- or multilayers of graphene
or $h$-BN are sandwiched between two ferromagnetic layers. The transport
direction is orthogonal to the plane of spacer layers. The key to feasibility
of such devices is the fact that the lattice constants of
graphene and $h$-BN match closely those of Co and Ni. Moreover, in the case 
of multilayer graphene the momentum selection criteria allow efficient
transport only for the minority-spin channel in the parallel configuration of the device. 
Very high magnetoresistance ratios have been predicted for multilayer 
graphene used as a spacer material \cite{Karpan07,Karpan08}. Meanwhile,
high-quality epitaxial monolayers of both graphene and $h$-BN on ferromagnetic 
transition metals have been grown experimentally using the chemical vapor deposition techniques 
\cite{Oshima97,Dedkov08a,Dedkov08b,Gruneis08,Rader09,Varykhalov08,Varykhalov09}. Successful growth 
of lower-quality multilayer graphene on polycrystalline Ni substrates
has also been reported \cite{Reina09}.  

Considerable progress has also been achieved in controlled manufacturing
of graphene nanostructures. The lithographic patterning allows to produce
graphene nanoribbons as narrow as $\sim$15~nm \cite{Han07}. Sub-10-nanometer
graphene nanoribbons have been synthesized using a variety of chemical approaches
starting from either graphite \cite{Wang08b,Li08} or carbon nanotubes \cite{Jiao09,Kosynkin09}. 
Well-ordered edges along a single crystallographic direction of graphene have been 
produced by means of chemical vapor deposition \cite{Campos-Delgado08},
annealing by Joule heating \cite{Jia09}, and anisotropic etching of graphene using
metallic nanoparticles \cite{Campos09}.

\subsection{Magnetic ordering at finite temperatures}

It has already been mentioned at the beginning that the Curie temperatures 
of ferromagnetic materials must be higher than the operation temperature of 
the device, which is typically supposed to be close to 300~K. When introducing the 
working principles of the proposed spintronic devices based on graphene edges, 
temperature-related limitations were not discussed. However, magnetic 
order in low-dimensional systems is particularly sensible to thermal fluctuations. 
In particular, the Mermin-Wagner 
theorem excludes long-range order in one-dimensional magnetic systems (such as
the magnetic graphene edges) at any finite temperature \cite{Mermin66}. 
The range of magnetic order is limited by the temperature-dependent spin correlation lengths 
$\xi^{\alpha}$ ($\alpha=x$, $y$, $z$) which define the decay law of the spin correlation
\begin{equation}
\la \hat{s}^{\alpha}_i \hat{s}^{\alpha}_{i+l} \ra = \la \hat{s}^\alpha_i \hat{s}^\alpha_{i} \ra {\rm exp}(-l/\xi^{\alpha}),
\end{equation}
where $\hat{s}^\alpha_i$ are the components of magnetic moment unit vector $\hat{\mathbf s}_i$
at site $i$. In principle, the spin correlation length imposes the limitations on the device dimensions. 
In order to establish this parameter one has to determine the energetics of the spin fluctuations 
contributing to the breakdown of the ordered ground-state configuration.

\begin{figure} 
\includegraphics[width=8.5cm]{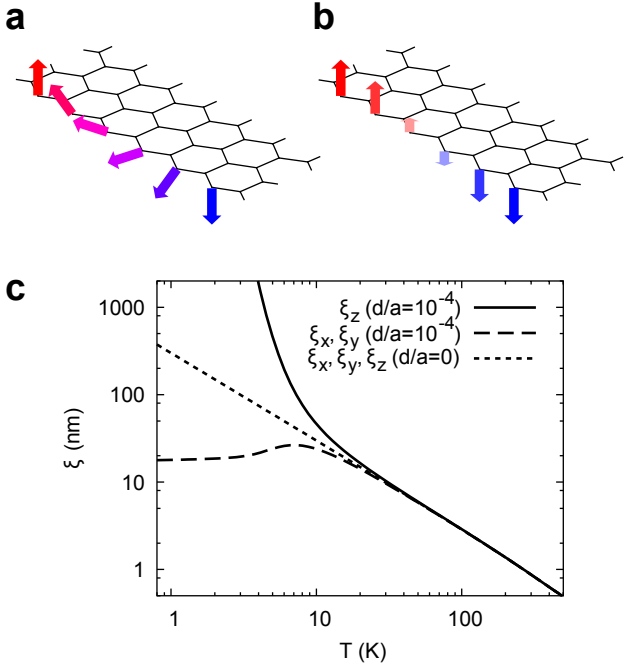}
\caption{\label{fig7} 
Schematic representation of the transverse (a) and the longitudinal (b) low-energy 
spin excitation at zigzag graphene edges. The magnetic moments are shown by arrows. 
Their directions are represented by orientation and color of the arrows, and the 
magnitude is shown by their lengths and color intensity. 
(c) Correlation lengths of magnetization vector components orthogonal 
($\xi_z$) and parallel ($\xi_x$, $\xi_y$) to the graphene plane as a function of temperature.
Reproduced from \cite{Yazyev08a}. Copyright 2008 by the American Physical Society.
}
\end{figure}

The energetics of the transverse and longitudinal spin excitations (Figs.~\ref{fig7}(a),(b)) 
have been explored using density-functional-theory calculations \cite{Yazyev08a}. 
The magnetic correlation parameters in the presence of spin-wave fluctuations, the dominant 
type of spin disorder in this case, were obtained with the help of one-dimensional
Heisenberg model Hamiltonian
\begin{equation}
    H = -a \sum_{i} \hat{\mathbf s}_i \hat{\mathbf s}_{i+1} - d \sum_{i} \hat{s}^z_i \hat{s}^z_{i+1},
\end{equation}
where the Heisenberg coupling $a=2\kappa/a_z^2$~=~105~meV corresponds to the spin-wave 
stiffness $\kappa=320$~meV~\AA$^2$ calculated from first principles. 
The estimated small anisotropy parameter $d/a \approx 10^{-4}$ originates from the weak spin-orbit 
interaction in carbon. This simple model Hamiltonian has known analytic solutions \cite{Joyce67}.
Figure~\ref{fig7}(c) shows the spin correlation lengths calculated for our particular case.
Above the crossover temperature $T_{\rm x}\approx 10$~K, weak magnetic anisotropy does not play any role and
the spin correlation length $\xi \propto T^{-1}$. However, below $T_{\rm x}$
the spin correlation length grows exponentially with decreasing temperature.
At $T=300$~K the spin correlation length $\xi \approx 1$~nm. 

From a practical point of view, this means that the dimensions of spintronic devices based 
on the magnetic zigzag edges of graphene and operating at normal temperature conditions 
are limited to several nanometers. At present, such dimensions are very difficult to achieve, 
which can be regarded as a pessimistic conclusion. Nevertheless, one has to keep in 
mind that the spin stiffness predicted for the magnetic graphene edges is still higher than 
the typical values for traditional magnetic materials. That is, graphene outperforms $d$-element
based magnetic materials, and there is a room for improvement. Achieving
control over the magnetic anisotropy $d/a$ could possibly raise the crossover
temperature $T_{\rm x}$ above 300~K and thus significantly extend $\xi$. Possible
approaches for reaching this goal include chemical functionalization of the edges with 
heavy-element functional groups or coupling graphene to a substrate.
  

\section{Magnetism in graphene and graphite}

\subsection{Radiation damage and defects in carbon materials}

Experimental observations of ferromagnetic ordering in irradiated graphite have
already been mentioned in the introductory part of this review. These results are particularly exciting
because of the fact that the induced magnetic ordering is stable at room temperature
and well above. Let us now try to understand the origin of magnetism in irradiated 
graphite. The present section covers the cases of both graphene and graphite which 
has a three-dimensional crystalline lattice composed of weakly coupled graphene 
layers. 

\begin{figure}[b]
\includegraphics[width=8.5cm]{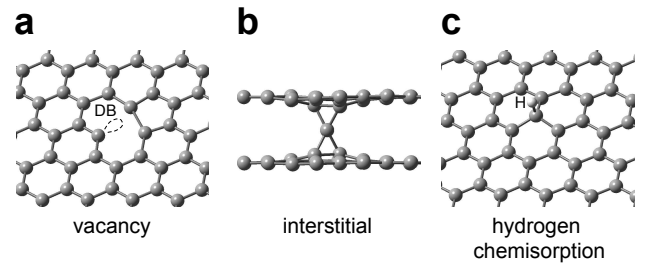}
\caption{\label{fig8} 
Atomic structures of primary point defects produced upon irradiating graphite
by high-energy protons: (a) single-atom vacancy, (b) interstitial bridging
the neighboring graphene layers, and (c) hydrogen chemisorption.
The dangling bond (DB) of the reconstructed vacancy defect and the chemisorbed
hydrogen atom (H) are labeled.   
}
\end{figure}

The basic picture of the radiation damage process in carbon materials is relatively 
simple. Irradiation of graphite with high-energy particles ({\it e.g.} protons) produces 
several types of point defects. In carbon materials the defects are created 
as a result of so-called ``knock-on collisions'' \cite{Banhart99,Krasheninnikov07}. 
This process involves the direct transfer of kinetic energy from the high-energy incident 
particles to the individual atoms in material's lattice. If the transferred energy is larger
than the displacement threshold $T_d$, the recoil atom may leave its equilibrium 
position leading to the formation of a pair of point defects -- a {\it vacancy} defect
and an {\it interstitial}. The structure of the vacancy defect in graphene and
graphite is shown in Figure~\ref{fig8}(a). In graphene the interstitial defects 
have a bridge structure \cite{Lehtinen03}, while in graphite the stable configuration
corresponds to a carbon atom trapped between the adjacent graphene layers \cite{Telling03,Li05}
as shown in Figures~\ref{fig8}(b).
The displacement threshold $T_d$ for carbon atoms in graphitic materials 
was found to be $\sim$20~eV in a number of studies \cite{Crespi96,Smith01,Yazyev07b,Zobelli07}. 
Creation of defects due to electron stopping, 
{\it i.e.} the process involving possible electronic excitations and ionization 
of individual atoms, is less important in carbon materials since electronic
excitations in metals are delocalized and quench instantly \cite{Banhart99,Krasheninnikov07}.

After slowing down, reactive particles may also produce chemisorption defects. In 
particular, protons are able to bind to individual carbon atoms in graphene lattice
resulting in their rehybridization into the $sp^3$-state (Fig.~\ref{fig8}(c)). 
Such defects are referred to as {\it hydrogen chemisorption} defects. From the point 
of view of one-orbital models that we use in our review, both vacancy and hydrogen 
chemisorption defects are equivalent. In both cases a defect removes one 
$p_z$-orbital from the $\pi$-system of graphene. In the first case, the $p_z$-orbital 
is eliminated together with the knocked-out carbon atom. The hydrogen chemisorption
does not remove the carbon atom from the crystalline lattice, but once rehybridized 
the atom is unable to contribute its $p_z$-orbital to the $\pi$-electron system. 
These two types of defects are further referred to as $p_z$-vacancies. 

The defects described above are the primary defects in the radiation-damage 
process. More complex defects can be produced at later stages of the process.
For instance, single-atom vacancies and interstitials may aggregate producing
extended defects. Complexes involving two or more different defects can also 
be formed upon irradiation. Examples are complexes of hydrogen with vacancies and interstitials
\cite{Lehtinen04}, and intimate Frenkel pairs \cite{Ewels03,Yazyev07b}. Radiation 
damage in graphitic materials may also produce the Stone-Wales defects \cite{Kaxiras88,Stone86}. 

\subsection{Defect-induced magnetism in graphene}

The single atom $p_z$-vacancies described above have a particularly profound effect 
on the electronic structure of ideal graphene. Let us consider a periodically repeated 
supercell of graphene composed of $2N$ ($N_A = N_B = N$) carbon atoms. Elimination of 
one atom from sublattice $A$ introduces a zero-energy state in the complementary sublattice 
($\alpha = N_B$; thus $\eta = 2N_B - ((N_A - 1) + N_B) = 1$). Such zero-energy states 
extending over large distances are called quasi-localized states since they show 
a power-law decay \cite{Pereira06,Huang09}. The quasi-localized states have 
been observed in a large number of scanning tunneling microscopy (STM) studies of graphite 
as triangular $\sqrt{3}\times\sqrt{3}R30^\circ$ superstructures extending over a few nanometers and
localized around point defects \cite{Mizes89,Kelly98,Ruffieux00}. 
For the single-defect model we have adopted, Lieb's theorem predicts a magnetic moment of 
$|(N_A - 1) + N_B | = 1\mu_{\rm B}$ per supercell, that is, the presence of a defect induces 
ferromagnetic ordering.

\begin{figure}[b]
\includegraphics[width=8.5cm]{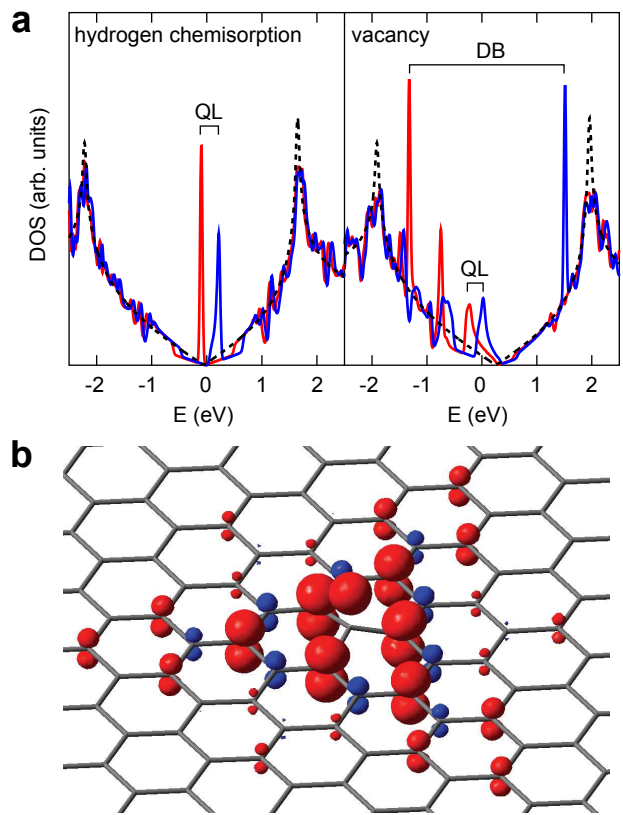}
\caption{\label{defectdos} 
(a) Spin-resolved density of states (DOS) for the vacancy and the hydrogen
chemisorption defect in graphene calculated from first principles. Red and blue curves
correspond to the majority and minority spins, respectively. Dashed curve shows
the reference density of states of the ideal graphene. Zero energy corresponds 
to the Fermi level. Exchange-split peaks which correspond to
quasi-localized (QL) and dangling-bond (DB) states are labeled.
(b) Isosurface representation of the spin-density distribution at the hydrogen
chemisorption defect obtained from first principles. Red and blue surfaces correspond to the majority- 
and minority-spin densities, respectively.  
}
\end{figure}

This result has been widely confirmed using both first-principles \cite{Duplock04,Lehtinen04,Yazyev07a} 
and mean-field Hubbard model \cite{Kumazaki07,Palacios08} calculations. Figure~\ref{defectdos}(a) 
shows the spin-resolved density-of-states plots for hydrogen
chemisorption and vacancy defects obtained using first-principles calculations \cite{Yazyev07a}. 
In the first case, the sharp peak close to the Fermi level corresponds to the quasi-localized 
state induced by the chemisorbed hydrogen atom. The peak is fully split by exchange 
and the system is characterized by a magnetic moment of 1$\mu_{\rm B}$ at any defect concentration.
The distribution of spin density around the defective site clearly shows 
a $\sqrt{3}\times\sqrt{3}R30^\circ$ superstructure (Fig.~\ref{defectdos}(b)).
The case of vacancy defect is somewhat more complicated. In addition to the 
quasi-localized state, there is also a localized non-bonding state due to the 
presence of a $\sigma$-symmetry dangling bond in this defect (Fig.~\ref{fig8}(a)). The dangling-bond
state shows a very strong exchange splitting and contributes 1$\mu_{\rm B}$ to the 
total magnetic moment of the defect (Fig.~\ref{defectdos}(a)). However, the
magnetic moment due to the quasi-localized state is partially suppressed in this case due
to the self-doping effect related to the structural reconstruction of the vacancy
\cite{Yazyev07a}. The overall magnetic moment per vacancy defect varies from 1.12$\mu_{\rm B}$
to 1.53$\mu_{\rm B}$ for defect concentrations ranging from 20\% to 0.5\%.

Magnetic moments due to dangling bonds can also be contributed by other 
types of defects, {\it e.g.} the bridge-configuration interstitial defect in graphene \cite{Lehtinen03}. 
However, one has to keep in mind that magnetic ordering due to only localized magnetic moments in graphene-based 
system is improbable at high temperatures. The Ruderman-Kittel-Kasuya-Yoshida interaction is expected to be
weak in this case due to the semi-metallic electronic structure of
graphene \cite{Vozmediano05,Dugaev06,Brey07,Saremi07}. On the other hand, magnetic ordering due to the quasi-localized states
can be considered as itinerant magnetism without excluding a possible contribution of 
dangling-bond magnetic moments to the net magnetic moment of a defective carbon system.

The system with one defect placed in a periodically repeated supercell 
is only a rough model of disordered graphene for two reasons. 
First, all defects are located in the same sublattice of the graphene layer. Second, 
the defects form an ordered periodic superlattice. A more realistic description 
of disorder can be achieved by constructing models with defects randomly distributed 
in a large enough supercell \cite{Yazyev08c}. Such models allow defects to occupy 
both sublattices at arbitrary concentrations and eliminate any short-range order 
in the spatial arrangement of defects. Larger supercells are needed for building
the disordered models which makes first-principles calculations impractical. However,
such system can still be treated using the mean-field Hubbard model calculations. 

\begin{figure}
\includegraphics[width=8.5cm]{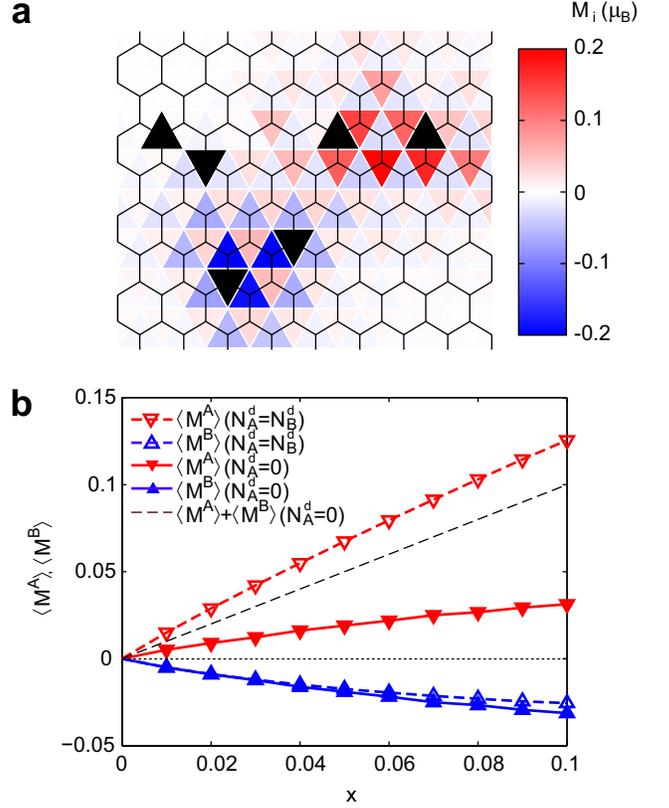}
\caption{\label{fig10}
(a) Distribution of local magnetic moments $M_i$ in a small region a system with random distribution of 
$p_z$-vacancies obtained thought the mean-field Hubbard model calculations.
Positions of defects belonging to sublattices $A$ and $B$ are marked by $\blacktriangle$ and 
$\blacktriangledown$, respectively.
(b) Average magnetic moments for the atoms in sublattice $A$ ($\triangledown$,$\blacktriangledown$) and sublattice $B$ ($\triangle$,$\blacktriangle$)
as a function of defect concentration $x$. The defects are either distributed equally 
between the two sublattices (solid curves/filled symbols) or belong to sublattice $B$ only (dashed curves/empty symbols). 
Net magnetic moments per carbon atom (dashed line) is shown for the case of defects distributed
over sublattice $B$ only.}
\end{figure}

Figure~\ref{fig10}(a) shows the distribution of spin density in a selected region 
of a large supercell randomly populated by $p_z$-vacancies. Defects in different sublattices
are shown as black triangles of different orientations. The resulting picture 
can be explained if one considers the following two arguments. 
First, from Lieb's theorem the total magnetic per supercell is $M = |N_A-N_B| = |N^d_B-N^d_A|$, 
where $N^d_A$ and $N^d_B$ are the numbers of defects created in sublattices $A$ and $B$, 
respectively. This means that electron spins populating the quasi-localized states 
in the same sublattice are oriented parallel to each other while the antiparallel
arrangement is realized when electron spins populate different sublattices.
This conclusion is qualitatively the same as the one we obtained when considering
finite graphene nanofragments. Second, quasi-localized states populating complementary 
sublattices interact with each other \cite{Kumazaki07,Palacios08}. The interaction lifts 
the degeneracy leading to weakly bonding and anti-bonding states. This provides 
another mechanism for escaping the instability associated with the presence of low-energy
electronic states.
The interaction strength  between two defects increases with decreasing distance 
between them. For very short distances, the gain in exchange energy does not compensate 
for the kinetic energy penalty due to the splitting. This leads to quenching of the defect-induced 
magnetic moments \cite{Boukhvalov08,Yazyev08c}. The competition between these effects is 
demonstrated in Figure~\ref{fig10}(a). 

More quantitative results are presented in Figure~\ref{fig10}(b) which shows the 
mean magnetic moment $\la M^A \ra$ and $\la M^B \ra$ per carbon atom in 
sublattice $A$ and sublattice $B$ as a function of defect concentration $x$ \cite{Yazyev08c}. 
The resulting values have been averaged over many random placements of defects in the simulation supercell. 
The plot refers to the situation of defects equally distributed over the two sublattices 
($N^d_A=N^d_B$) and to the situation when defects belong to sublattice $B$ only 
($N^d_A=0$). In the first case, the magnetic moments in the two sublattices
compensate each other. The overall magnetic ordering is of {\it antiferromagnetic} character.
When defects populate only one sublattice the system exhibits {\it ferromagnetic} ordering. 
The net magnetic moment per carbon atom $\la M \ra = (\la M^A \ra + \la M^B \ra)/2 = x/2$ 
scales linearly with the defect concentration. Both numerical results are in full 
agreement with Lieb's theorem.

\subsection{Magnetism in graphite and multilayer graphene}

\begin{figure}[b]
\includegraphics[width=8.5cm]{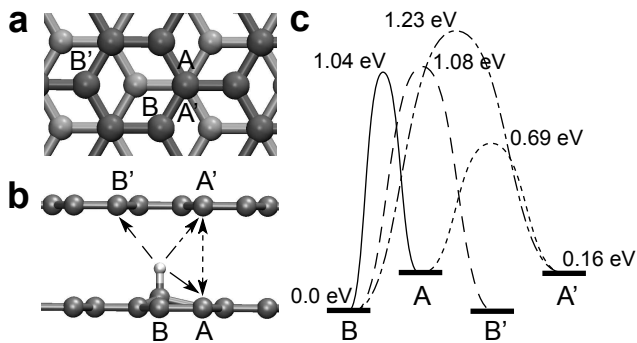}
\caption{\label{graphite} 
(a) Inequivalent carbon atoms ($A$ and $B$) in $ABA$ graphite.
(b) Possible pathways for the diffusion of chemisorbed hydrogen
in graphite. (c) Schematic representation of the potential energy surface for the in-plane diffusion of
hydrogen in graphite showing the relative energies for the local minima and transition states.
Reproduced from \cite{Yazyev08c}. Copyright 2008 by the American Physical Society.
}
\end{figure}

Experimental observations of ferromagnetism in irradiated graphite point to the 
conclusion that sublattices of individual graphene layers in bulk graphite are 
populated by defects differently: that is, there must be a mechanism which 
makes the sublattices of graphene inequivalent. Such an intrinsic discriminating 
mechanism was ascribed to the stacking order of graphite layers in bulk graphite 
\cite{Yazyev08c}. The lowest-energy $ABA$ stacking order of individual graphene 
sheets in graphite breaks the equivalence of the two sublattices as shown in 
Figure~\ref{graphite}(a). In fact, only local $ABA$ stacking order is required to discriminate 
between the two sublattices of the middle sheet. 
The mechanism can be demonstrated for the case of hydrogen chemisorption
defects. First-principles calculations show that the configuration which involves 
hydrogen chemisorbed on sublattice $B$ is 0.16~eV lower in energy than on sublattice $A$ 
(see Figs.~\ref{graphite}(b) and \ref{graphite}(c)). This energy difference is sufficient to trigger a considerable 
difference in equilibrium populations of the two sublattices. The energy barrier for 
the hopping of chemisorbed hydrogen atoms is relatively small ($\sim$1~eV)
to allow for thermally activated diffusion at mild temperatures.  

Similar discriminating mechanisms may also exist for the other types of defects created by
irradiation, {\it e.g.} for vacancies. Cross-sections for momentum transfer during 
knock-on collisions with high-energy incident particles are likely to be very similar 
for both $A$ and $B$ carbon atoms in graphite. However, the stacking order may have a strong 
influence on the recombination of interstitial and vacancy defects close to equilibrium
conditions. It was also shown that instantaneous recombination of low-energy recoil atoms 
in graphite is significantly more probable for atoms in position $A$ \cite{Yazyev07b}. 
That is, more vacancies in sublattice $B$ are left assuming an equal number of knock-on
collisions involving the atoms of both types. These results allow us to conclude that the most 
probable physical picture of magnetic order in irradiated graphite is {\it ferrimagnetism}. 
The magnetic moment induced by defects in sublattice $A$ is larger than the one 
induced in sublattice $B$.

It is worth mentioning other possible scenarios for the onset of magnetism in 
graphene-related materials and nanostructures. It was shown that local negative 
Gauss curvatures in graphene layers also lead to localized magnetic moments \cite{Park03}.
Coupling between the two graphene layers in a biased graphene bilayer introduces 
a pair of sharp van~Hove singularities close to the Fermi level. It was shown theoretically
that in a range of conditions biased bilayer graphene satisfies the Stoner criterion
leading to a low-density ferromagnetic phase \cite{Castro08}.   

The question of magnetic ordering in defective graphene and graphite at finite 
temperatures remains largely unaddressed. Similarly to the one-dimensional  
system discussed above, an isotropic two-dimensional system cannot develop 
long-range magnetic ordering at any finite temperature \cite{Mermin66}. 
However, the introduction of a small magnetic anisotropy $d/a \sim 10^{-3}$
already leads to very high transition temperatures \cite{Barzola-Quiquia07}.
Weak magnetic coupling between the individual layers in graphite also produces 
a pronounced effect on the magnetic transition temperature \cite{Pisani08}.

\section{Conclusions and perspectives}

The review illustrated a rich variety of magnetism scenarios in graphene nanostructures 
and explained them in terms of simple physical models. Beyond these theoretical 
considerations the field of carbon-based magnetism faces a number of challenges. 
The most important problems are related to the experimental side of 
the field. In particular, the physics of magnetic graphene edges has already 
attracted a large number of computational and theoretical researchers. However, 
no direct experimental evidence has been reported at the time this review 
was written. Further progress in this field will also require novel manufacturing 
techniques which would allow control of the edge configuration with truly atomic
precision. The area of defect-induced magnetism in graphite demands detailed 
studies of defects produced upon irradiation and their role in the onset of ferromagnetic ordering. 
The limits of saturation magnetization and Curie temperature in irradiated graphite 
have still to be established. On the theory side of this field, an understanding 
of magnetic phase transitions in graphene materials and nanostructures has to 
be developed. Other important directions of theoretical research include spin 
transport and magnetic anisotropy of carbon-based systems. 

\section{Acknowledgments}

I would like to thank J. Fern\'andez-Rossier, Y.-W. Son and D.~Strubbe for critical reading of the manuscript.
This work was supported by the Swiss National Science Foundation (grant No. PBELP2-123086).

\bibliography{tutorial}

\end{document}